\documentstyle[12pt]{article}
\newcommand{\ie}{{\it i.e.}}
\newcommand{\eg}{{\it e.g.}}

\newcommand{\gsim}{\buildrel > \over {_\sim}}
\newcommand{\lsim}{\buildrel < \over {_\sim}}
\newcommand{\order}{{\cal O}}
\newcommand{\ccbar}{c\bar{c}}

\newcommand{\qqbar}{q\bar{q}}
\newcommand{\jp}{J/\psi}
\newcommand{\as}{$\alpha_s$}
\newcommand{\BRone}{{\rm Br}(\chi_1 \to \psi \gamma)}

\newcommand{\rhII}{{\varrho}_{11}}
\newcommand{\rhOO}{{\varrho}_{00}}
\newcommand{\kbt}{{\bf k}_{1\perp}}
\newcommand{\ktt}{{\bf k}_{2\perp}}
\newcommand{\kbz}{k_{1z}}
\newcommand{\ktz}{k_{2z}}
\newcommand{\RS}{|R_S(0)|^2}
\newcommand{\RP}{|R'_P(0)|^2}
\newcommand{\als}{\alpha_s}
\newcommand{\lra}{\leftrightarrow}
\newcommand{\half}{\frac{1}{2}}

\newcommand{\PLB}[3]{\mbox{}Phys. Lett. {\bf B{#1}}, {#2} ({#3})}
\newcommand{\NPB}[3]{\mbox{}Nucl. Phys. {\bf B{#1}}, {#2} ({#3})}
\newcommand{\PRL}[3]{\mbox{}Phys. Rev. Lett. {\bf {#1}}, {#2} ({#3})}
\newcommand{\PRD}[3]{\mbox{}Phys. Rev. {\bf D{#1}}, {#2} ({#3})}
\newcommand{\ZPC}[3]{\mbox{}Z. Phys. {\bf C{#1}}, {#2} ({#3})}

\newcommand{\etal}{{\em et al.}}

\begin{document}

\setlength{\baselineskip}{7mm}

\thispagestyle{empty}
\begin{flushright}
   \vbox{\baselineskip 12.5pt plus 1pt minus 1pt
         SLAC-PUB-6637 \\
         HU-TFT-94-29 \\
         August 1994 \\
         (T/E)
             }
\end{flushright}

\renewcommand{\thefootnote}{\fnsymbol{footnote}}
\bigskip
\begin{center}
{\Large \bf Hadroproduction and Polarization of Charmonium}

\vskip 1\baselineskip

M. V\"anttinen\footnote
{Work supported by the Academy of Finland
under project number 8579.}$^a$ and P. Hoyer$^*$\footnote
{Address after 1
September 1994: Nordita, Copenhagen.}$^b$ \\
{\normalsize \em $^a$Research Institute
for Theoretical Physics and} \\ {\normalsize \em $^b$Department of
Physics} \\ {\normalsize  \em University of Helsinki, Finland}

\vskip 1\baselineskip

  S. J. Brodsky\mbox{\scriptsize $^{\ddag}$} and
W.-K. Tang\footnote{Work supported by the Department
 of Energy, contract DE-AC03-76SF00515.}
 \\ {\normalsize \em
  Stanford Linear Accelerator Center} \\ {\normalsize \em Stanford
  University, Stanford, CA 94309}

\end{center}

\medskip

\renewcommand{\thefootnote}{\arabic{footnote}}
\setcounter{footnote}{0}

\begin{center}
{\bf ABSTRACT}
\end{center}

\vbox{\baselineskip 14pt
\noindent
In the limit of heavy quark mass, the production cross section and
polarization of quarkonia can be calculated
in perturbative QCD. We study the
$p_\perp$-averaged production of charmonium states in $\pi N$ collisions
at fixed target energies. The data on the relative production
rates of $\jp$ and $\chi_J$ is found to disagree with leading twist QCD.
The polarization of the $\jp$ indicates that the discrepancy is
not due to poorly known parton distributions
nor to the size of higher order effects ($K$-factors).
Rather, the disagreement
suggests important higher twist corrections, as has
been surmised earlier
from the nuclear target $A$-dependence of the production cross section.
}

\bigskip
\begin{center}
{\it Submitted to Physical Review D.}
\end{center}
\newpage
\section{Introduction}

Quarkonium bound states formed by heavy quark-antiquark pairs are small
nonrelativistic systems, whose production and decay
properties are expected to
be governed by perturbative QCD.
The extensive data available on the inclusive decays of
many charmonium ($\ccbar$) and bottomonium
($b\bar b$) states has been compared
with detailed perturbative calculations.
The overall agreement between theory
and experiment is reasonably good, taking into account the moderate mass
scale
\cite{Kwong,Kopke,Schuler,Braaten}.
This work has led to self-consistent
values of the size of the quarkonium wave function near the origin.

Assuming, then, that we understand the decay of the quarkonium states to
perturbative gluon and light-quark final states,
we can turn the reaction around
and consider the photoproduction and hadroproduction of quarkonia.
Thus quarkonium production becomes a probe of the production mechanism
of color-singlet heavy quark pairs. This is analogous to the dynamics of
lepton pair hadroproduction, where the main production mechanism
(at lowest order) has been identified as the
Drell-Yan hard fusion subprocess
$\qqbar \to \gamma^*$. Quarkonium production
can offer new insights into gluon
fusion mechanisms; for example, the $\jp$ and $\chi_1$
couple to states with more than two light partons, such as $ggg$ or
$\qqbar g$. At leading twist, \ie, to leading order in $1/m_Q$,
quarkonium production proceeds through the
collision of only two partons, one
from the projectile and one from the target. Hence an extra gluon or quark
must be emitted in the leading twist production of $\jp$ and $\chi_1$.
However, at large values of the quarkonium
momentum fraction $x_F,$ it becomes
advantageous for two or more collinear
partons from either the projectile or target
to participate in the reaction.
Such processes  are higher twist, since their
rate is suppressed relative to
ordinary fusion reactions by powers of $\Lambda_{QCD}/m_Q$ where
$\Lambda_{QCD}$ is the characteristic transverse momentum in the incident
hadron wavefunction.
Nevertheless, despite
the extra powers of $1/m_Q$, the multiparton
processes can become dominant at
$(1-x_F) <{\cal O}(\Lambda_{QCD}^2/m_Q^2)$
since they are efficient in converting the incident hadron momentum into
high $x_F$ quarkonia \cite{BHMT}.
\eject

In leading twist QCD
the production of the
$\jp$ at low transverse momentum occurs both `directly'
from the gluon fusion subprocess $gg \to \jp+g$
and indirectly
via the production of $\chi_1$ and $\chi_2$ states\footnote{At high
transverse momentum, one also has to take into account production
through quark and gluon fragmentation \cite{BY}.}.
These states
have
sizable decay branching fractions $\chi_{1,2} \to \jp+\gamma$
of 27\% and 13\%,
respectively.
In spite of its relatively small branching ratio, the $\chi_2$
state is expected to
give an important contribution to the total yield of $\jp$'s
at leading twist, since $gg \to \chi_2$ is of lower order in $\alpha_s$
compared to the competing processes. Early comparisons
\cite{Carlson,Kuhn,Chang,BaierZPC19}
with the total $\jp$ cross section data
indicated rough agreement with the  model predictions.
Nevertheless, the
cross sections for direct $\jp$ and $\chi_1$ production were predicted
\cite{BargerPRD31} to be too low compared to the data
\cite{Clark,Kourkoumelis,WA11,E673,Binon}.

More recent E705 and E672
data \cite{AntoniazziPRL70,Zieminski} on the
production fractions of the various charmonium
states have confirmed that there
is a clear discrepancy with the leading twist
QCD prediction. The leading twist
calculations which we present in this paper
show that the predicted ratio of direct
$\jp$ production in
$\pi N$ collisions
compared to the
$\chi_2$ production is too low by a factor
of about 3.
In addition, the ratio of
$\chi_1$ production to $\chi_2$ production
is too low by a factor of 10.
A similar conclusion has been
reached in \cite{Schuler},
where possible explanations in terms of uncertainties
in the partonic cross sections
(very different $K$-factors for the various
processes) or unconventional
pion
parton distributions
are discussed. Less data
is available for proton-induced charmonium production, but a discrepancy
between leading twist QCD
and experiment appears likely also in that case.

The wealth of data from the NA3 experiment at CERN \cite{Badier}
and the
Chicago-Iowa-Princeton \cite {Biino}
and E537 experiments \cite {Akerlof}
at FermiLab
on the angular distribution of the muons in the
decay $\jp \to \mu^+\mu^-$ provides an even more sensitive
discriminant of
different production
mechanisms
\cite{Ioffe,Berger,ArgyresPRD26,ArgyresNPB234,BaierNPB201,%
BaierNPB218,Korner}.

The polarization of the $\ccbar$,
and hence that of the charmonium bound state \cite{Berger},
can at leading twist be calculated from perturbative QCD.
Furthermore, in the heavy quark
limit, the radiative transition $\chi_J \to \jp+\gamma$
preserves the quark
spins,
\ie, it is an electric dipole transition.
Hence the polarization
also of indirectly produced $\jp$'s
can be calculated. We find that even if the
relative production rates
of the $\jp,\  \chi_1$ and $\chi_2$ are adjusted
(using $K$-factors) to agree with the data,
the $\jp$ polarization data is still not reproduced.

We shall argue that a
possible
explanation for the underestimate of the $\jp$
and $\chi_1$ cross sections is that more than one parton from either the
projectile or target participates in the collision, so that
no additional gluon
needs to be emitted.
Similar higher twist effects are known to become important
at high $x_F$ in lepton pair
production \cite{BerBro,BBKM,Eskola,Heinrich}.
In Ref. \cite {BBKM} it is shown that higher twist contributions can
explain the large azimuthal $\cos \phi$ and $\cos 2 \phi$ correlations
seen in the $\pi N \to \mu^+ \mu^-$ data.
There are also previous indications
from the non-factorizing anomalous nuclear target dependence of the $\jp$
cross section \cite{HVS,Badier,AldeDY,Katsanevas,AldeJP}
that higher twist
effects are considerably larger in $\jp$ production
than in lepton pair production, and that they persist down to low $x_F$.

\section{Production rates of $\psi$ and $\chi_J$ states at leading twist}

In this section we calculate $\jp$ production
in $\pi N$ interactions at leading
twist and to
lowest order\footnote{Thus we do not include subprocesses like
$qg\rightarrow \chi_2 q$ (which is subleading to $gg \rightarrow
\chi_2)$.} in \as.
Higher order corrections in \as\ and relativistic
corrections to the charmonium bound states are unlikely to change our
qualitative conclusions at moderate $x_F$.
Contributions from direct $\jp$
production, as well as from indirect production via $\chi_1$ and $\chi_2$
decays, are included. Due to the small branching fraction
$\chi_0 \to \jp+\gamma$ of 0.7\%, the contribution from $\chi_0$ to $\jp$
production is expected (and observed) to be negligible. Decays from the
radially excited $2^3S_1$  state, $\psi' \to \jp + X$, contribute
to  the total $\jp$ rate at the few per cent level
and also will be ignored here.

Since the $\psi'$ is formed directly,
its production allows an important cross
check on the use of charmonium states
to study the production mechanism. At
high energies, the charmonium bound
state forms long after the production of
the compact $\ccbar$ pair (the formation time
$\tau_{\rm form} \sim 2E_{\rm lab}/\Delta M^2$).
Thus the ratio of $\psi'$ to direct
$\jp$ production can depend only on the relative magnitude of their wave
functions at the origin. More precisely (see, \eg,~\cite{Schuler}),
\begin{equation}
\frac{\sigma(\psi')}{\sigma_{\rm dir}(\jp)}
\simeq \frac{\Gamma(\psi'\to e^+e^-)}
  {\Gamma(\jp\to e^+e^-)} \frac{M_{\jp}^3}
  {M_{\psi'}^3} \simeq 0.24 \pm 0.03
  \label{psiprime}
\end{equation}
where $\sigma_{\rm dir}(\jp)$ is the cross
section for direct production of the
$\jp$. The
ratio (\ref{psiprime}) should hold for all beams and targets,
independent of
the size of the higher twist corrections
in producing the pointlike $\ccbar$
state. The energy should be large enough
for the bound state to form outside the
target. The available data is indeed compatible with (\ref{psiprime}). In
particular, the E705 value \cite{AntoniazziPRL70} is about $0.24$ (see
Table 1). The anomalous nuclear target
$A$-dependence observed for the $\jp$
is also seen for the $\psi'$ \cite{AldeJP},
so that the ratio (\ref{psiprime})
is indeed independent of $A$.

\begin{table}[htb]
\begin{center}
\begin{tabular}{|c|c|c|c|}
  \hline
  & $\sigma(\psi')$ [nb] & $\sigma_{\rm dir}(\jp)$[nb]
    & $\sigma(\psi')/\sigma_{\rm dir}(\jp)$ \\ \hline
  $\pi^+$ & $22\pm 5$ & $97\pm 14$  & $0.23\pm 0.07$ \\ \hline
  $\pi^-$ & $25\pm 4$ & $102\pm 14$ & $0.25\pm 0.05$ \\ \hline
  $p$     & $20\pm 3$ & $89\pm 12$  & $0.23\pm 0.05$ \\ \hline
\end{tabular}
\end{center}
\caption{Production cross sections for $\psi'$,  direct
$\jp$ and their ratio in $\pi^+ N$, $\pi^- N$ and $p N$ collisions.
The data are from Ref. \protect\cite{AntoniazziPRL70}.}
\end{table}

The $\pi N \to \chi_2+X$ production cross section
to lowest order and twist is
\begin{equation}
  \sigma(\pi N \to \chi_2+X;\ x_F>0)
    = \tau \int_{\sqrt{\tau}}^1 \frac{dx_1}{x_1}
    F_{g/\pi}(x_1) F_{g/N}(\tau/x_1) \sigma_0(gg\to \chi_2)
    \label{sigchi2}
\end{equation}
where
$\tau=M_{\chi_2}^2/s$ and the quantity
$\sigma_0(gg\to \chi_2) = 16 \pi^2 \als^2 \RP / M_{\chi_2}^7$
\cite{BaierZPC19}. We restrict the $\chi_2$ momentum range
to the forward CM
hemisphere $(x_F>0)$ in accordance with the available data.

The direct $\pi N \to \jp+X$ cross section is similarly given by
\begin{eqnarray}
  \sigma(\pi N \to \jp+X;\ x_F>0) & = & \int_\tau^1 dx_1
  \int_{\tau/x_1}^1 dx_2
  \int_{\widehat t_{\rm min}}^0 d\widehat t F_{g/\pi}(x_1) F_{g/N}(x_2)
  \nonumber \\
  & & \times \frac{d\sigma}{d\widehat t}(gg\to \jp+g) \label{sigjp}
\end{eqnarray}
where $\widehat
t$ is the invariant momentum transfer in the subprocess, and
\begin{equation}
  \widehat t_{\rm min} = {\rm max} \left(
\frac{x_2 M_{\jp}^2 - x_1
\widehat s}{x_1 + x_2}, M_{\jp}^2 - \widehat s \right).
\end{equation}
Eq. (\ref{sigjp}) also applies to the $\pi N \to
\chi_1+X$ reaction, in which
case a sum over the relevant subprocesses $gg
\to \chi_1g$, $g\bar{q} \to \chi_1
\bar{q}$,
$gq \to \chi_1 q$
and $\qqbar \to \chi_1 g$ is necessary. The differential cross sections
$d\sigma/d\widehat t$ for all subprocesses are given in
\cite{BaierZPC19,GastmansWu}.
In Table 2 we compare the $\chi_2$ production
cross section and the relative rates of direct
$\jp$ and $\chi_1$ production
at $E_{lab}=300$ GeV
with the data of E705 and WA11 on $\pi^-N$ collisions
at $E_{lab}=300$ GeV
and 185 GeV \cite{WA11,AntoniazziPRL70}.
We use the parton distributions of Ref. \cite{OwensPRD30,OwensPLB266}
evaluated at $Q^2=M^2$, where $M$ is the mass of the charmonium state
in question. We take $\alpha_s=0.26$ for all states and use
$|R_{S}(0)|^2=0.7 \; {\rm GeV}^3$, $|R'_P(0)/M|^2=0.006 \; {\rm GeV}^3$
\cite{wavefunctions}.

\begin{table}[htb]
\begin{center}
\begin{tabular}{|c|c|c|c|}
  \hline
  & $\sigma(\chi_2)$ [nb] & $\sigma_{\rm dir}(\jp)/\sigma(\chi_2)$
    & $\sigma(\chi_1)/\sigma(\chi_2)$ \\ \hline
  Experiment & $188 \pm 30\pm 21$  & $0.54 \pm 0.11\pm 0.10$ & $0.70 \pm
  0.15\pm 0.12$
   \\ \hline
  Theory     & 78 & 0.17 & 0.067 \\ \hline
\end{tabular}
\end{center}
\caption{Production cross sections for
$\chi_1$, $\chi_2$ and directly produced
$\jp$ in $\pi^- N$ collisions at 300 GeV. The data from Ref.
\protect\cite{WA11,AntoniazziPRL70}
include measurements at 185 and 300 GeV.}
\end{table}

The $\chi_2$ production rate in QCD agrees with the data within a
`$K$-factor' of order $2$ to $3$.
This is within the theoretical uncertainties
arising from the $\jp$ and $\chi$
wavefunctions, higher order corrections,
parton distributions,
and the renormalization scale. A similar factor is found between the
lowest-order QCD calculation and the data on lepton pair production
\cite{BadierDY,Conway}. On the
other hand,
Table 2 shows a considerable discrepancy between the calculated and
measured relative production rates of direct $\jp$ and $\chi_1$
compared to
$\chi_2$ production. A priori we would
expect the $K$-factors to be roughly
similar for all three processes.
It should be noted that there is a kinematic
region in the $\jp$ and $\chi_1$
processes where the emitted parton is soft (in
the rest frame of the charmonium),
and where perturbation theory could fail.
However, the contribution from this region is numerically not important
(there is actually no infrared divergence).
Hence one cannot hope to boost
the cross section significantly by multiplying
the soft parton contribution
by any reasonable factor. Moreover, the same soft parton region exists in
charmonium decays, where analogous disagreements with data are absent.
It should also be noted that the contribution
to $\chi_1$ production from the $ \qqbar \to \chi_1+g$ subprocess
is apparently singular at
$\widehat s = M_{\chi_1}^2$ due to a
breakdown of the non-relativistic approximation
of the bound state
\cite{BargerPRD31}.
The divergence is cancelled once one takes
into account higher Fock states \cite{Braaten}.
In
agreement with Ref. \cite{Schuler,Doncheski140}
we find that the cross section
is insensitive to the value of the cutoff parameter
excluding the soft gluon
region.

We conclude that leading twist QCD appears to be in
conflict with the observed rate of
direct $\jp$ and $\chi_1$ production.
Although in Table 2 we only compared our
calculation with the E705 and WA11 $\pi^- N$ data, this comparison is
representative of the overall situation
(for a recent comprehensive review
see \cite{Schuler}).

\section{Polarization of the $\jp$}

The polarization of the $\jp$ is determined
by the angular distribution of its
decay muons in the $\jp$ rest frame.
By rotational symmetry and parity, the
angular distribution of massless muons,
integrated over the azimuthal angle, has
the form
\begin{equation} \frac{d\sigma}{d\cos\theta} \propto 1 + \lambda \cos^2
\theta \label{lambda}
\end{equation}
where we take $\theta$ to be the
angle between the $\mu^+$ and the projectile
direction (\ie, we use the Gottfried--Jackson frame).
The parameter $\lambda$
 can
be calculated from the $\ccbar$ production
amplitude and the electric dipole
approximation of radiative $\chi$ decays. Earlier calculations of the
polarization in hadroproduction
\cite{Ioffe,ArgyresPRD26,ArgyresNPB234} were
based on general effective couplings of the quarkonia and partons rather
than the perturbative-QCD matrix elements which we shall use.

The electric dipole approximation of the radiative decay
$\chi_J \to \psi \gamma$ is exact
in the heavy quark limit, \ie, when terms of
$\order(E_\gamma/m_c)$ are neglected. As a consequence, the heavy quark
spins are conserved in the decay,
while the orbital angular momentum changes.
This spin conservation may also be derived from Heavy Quark Symmetry
\cite{HQET}. The validity of the
electric dipole approximation for $\chi_J$
radiative decays has been verified experimentally \cite{dipole}.
\eject

The amplitude for the $\chi_2$ production subprocess
    $g(\mu_1)g(\mu_2) \to \ccbar \to \chi_2(J_z)$
is, following the notation of Ref. \cite{Guberina},
\begin{eqnarray}
  A(J_z=\pm 2) & = & 4 \als R'_P(0) \sqrt{\frac{\pi}{M^3}}
    \; e^{\mp 2i\phi} \nonumber \\
    & & \times [ 1 \mp (\mu_2-\mu_1) \cos\vartheta
    - \mu_1\mu_2 \cos^2 \vartheta
    - \delta_{ \mu_1\mu_2} \sin^2 \vartheta ] \\
  A(J_z=\pm 1) & = & 4 \als R'_P(0) \sqrt{\frac{\pi}{M^3}}
    \sin\vartheta \; e^{\mp i\phi} \nonumber \\
    & & \times [ \mu_1 - \mu_2
    \mp 2 \mu_1\mu_2 \cos\vartheta
    \pm 2 \delta_{ \mu_1\mu_2} \cos\vartheta ] \\
  A(J_z=0) & = & 4 \als R'_P(0) \sqrt{\frac{\pi}{M^3}}
    \sqrt{6} \; \delta_{ \mu_1, -\mu_2} \sin^2 \vartheta
\end{eqnarray}
where
$\vartheta,\phi$ are the polar and azimuthal angles of the beam gluon in
the Gottfried--Jackson frame; $\vartheta=0$
if the transverse momenta of the
incoming gluons are neglected.
In this case, as expected for physical,
transversely polarized gluons with
$\mu_{1,2} = \pm 1$, the amplitude for $\chi_2$ production with
$J_z = \mu_1-\mu_2 = \pm 1$ vanishes.
Surprisingly, the amplitude for $J_z=0$
also vanishes when $\vartheta=0$.
Hence the $\chi_2$ is at lowest order
produced only with $J_z=\pm 2$. In this
polarization state the spin and orbital angular momenta
of its constituent charm quarks are aligned,
$S_z = L_z = \pm 1$. Since $S_z$ is
conserved in the radiative decay $\chi_2 \to \jp+\gamma$, it follows that
$J_z(\jp)=S_z= \pm 1$ ($L=0$ for the $\jp$). Thus the $\jp$'s produced
via $\chi_2$ decay are transversely polarized, \ie, $\lambda=1$ in
the angular distribution
(\ref{lambda}). This result is exact if both
the photon recoil and the intrinsic
transverse momenta of the incoming partons are neglected.
Smearing of the beam
parton's transverse momentum distribution by a Gaussian function
$\exp \left[ -(k_\perp/500 \; {\rm MeV})^2 \right] $ would
reduce   $\lambda$ to  $\simeq 0.85$.

{}From the $gg\to\jp+g$ amplitude we find for  direct
$\jp$ production, $\pi N \to \jp+X \to \mu^+\mu^-+X$,
\begin{eqnarray}
  \frac{1}{B_{\mu\mu}}\frac{d\sigma}{dx_F d\cos\theta} =
  \frac{3}{64\pi}
  \int \frac{dx_1 dx_2}{(x_1 + x_2)
  \widehat{s}} F_{g/\pi}(x_1) F_{g/N}(x_2)
  \nonumber \\ \times
  \left[ \rhII + \rhOO + (\rhII - \rhOO) \cos^2 \theta \right]
  \label{direct_distribution}
\end{eqnarray}
where $B_{\mu\mu}$ is the $\jp\to\mu^+\mu^-$ branching fraction,
$x_F=2p^z_\psi/\sqrt{s}$ is the longitudinal-momentum
fraction of the $\jp$,
and $\theta$ is the muon decay angle of Eq.
(\ref{lambda}). The density matrix
elements $\rhII, \rhOO$ are given in the Appendix.
\goodbreak

For the $\pi N\to \chi_1+X \to \jp+\gamma+X \to \mu^+\mu^-+\gamma+X$
production
process we obtain  similarly
\begin{eqnarray}
  \frac{1}{B_{\mu\mu}}\frac{d\sigma}{dx_F d\cos\theta}
  & = & \frac{3}{128\pi} \BRone
  \sum_{ij} \int \frac{dx_1 dx_2}{(x_1 + x_2) \widehat{s}}
  F_{i/\pi}(x_1) F_{j/N}(x_2)
  \nonumber \\ & & \times
\left[ \rhOO^{ij} + 3\rhII^{ij} + (\rhOO^{ij} - \rhII^{ij}) \cos^2 \theta
  \right], \label{chi1_distribution}
\end{eqnarray}
where the density matrix elements for $ij$ =
$gg$, $gq$, $g\bar{q}$ and $\qqbar$
scattering
are again given in the Appendix.

In Fig. 1a we show the predicted values of the parameter $\lambda$ of Eq.
(\ref{lambda}) in the
Gottfried--Jackson
frame as a function of $x_F$, for the direct $\jp$ and the
$\chi_{1,2} \to
\jp+\gamma$ processes separately.
Direct $\jp$ production gives $\lambda \simeq 0.25$
in the moderate $x_F$ region,
whereas production via $\chi_1$ results in
$\lambda \simeq -0.15$. The dashed
lines indicate the effect of a Gaussian
smearing in the transverse momentum of
the beam partons.

The $\lambda(x_F)$-distribution obtained
when both the direct and indirect
$\jp$ production processes are taken into account
is shown in Fig. 1b and
compared with the
Chicago--Iowa--Princeton \cite{Biino} and E537 \cite{Akerlof} data.
Our QCD calculation gives $\lambda \simeq 0.5$ for $x_F \lsim
0.6$, significantly different from the measured value $\lambda \simeq 0$.
The E537 data gives $\lambda=0.028 \pm 0.004$ for $x_F>0$,
to be compared with our calculated value
$\lambda=0.50$ in the same range.

The discrepancies between the calculated and measured values of $\lambda$
are
one further indication that the standard leading twist processes
considered here are not adequate for explaining charmonium production.
The $\jp$ polarization is particularly sensitive to the production
mechanisms and allows us to make further conclusions on
the origin of the disagreements, including the above discrepancies in
the relative production cross sections of $\jp$, $\chi_{1}$ and
$\chi_{2}$. If these discrepancies arise from an incorrect
relative normalization of the various
subprocess contributions (\eg, due to
higher order effects), then we would
expect the $\jp$ polarization to agree
with data when the relative rates of
the subprocesses are adjusted according to
the measured cross sections of direct $\jp$, $\chi_1$ and $\chi_2$
production\footnote{In the case of
Drell-Yan virtual photon production, it is
known that higher order
corrections do not change the $\gamma^*$ polarization significantly
\cite{Chiappetta}, which makes it
plausible to represent these corrections by a
simple multiplicative factor
that
does not affect the polarization of the photon.}. The
lower
curve in Fig. 1b shows the effect of multiplying the partial
$\jp$ cross sections with the required
$K$-factors. The smearing effect is insignificant as shown by the
dashed curve. The $\lambda$ parameter is
still predicted incorrectly over most of the $x_F$ range.

A similar conclusion is reached
(within somewhat larger experimental errors) if
we compare our calculated value for the polarization of direct $\jp$
production, shown in Fig. 1a,
with the measured value of $\lambda$ for $\psi'$ production. In analogy
to Eq. (\ref{psiprime}), the $\psi'$ polarization data
should agree with the polarization of directly produced
$\jp$'s, regardless of the production mechanism.  Based on the angular
distribution of the muons from $\psi'
\to \mu^+\mu^-$ decays in 253 GeV $\pi^-W$
collisions, Ref. \cite{Heinrich} quotes
$\lambda_{\psi'} = 0.02 \pm 0.14$ for
$x_F>0.25$, appreciably lower than our QCD values for direct $\jp$'s
shown in Fig. 1a.

\section{Discussion}

We have seen that the $\jp$ and $\chi_1$
hadroproduction cross sections in
leading twist QCD are at considerable
variance with the data, while the $\chi_2$
cross section agrees with measurements within a reasonable $K$-factor of
2 to 3. On the other hand, the inclusive decays of the charmonium states
based on  the minimal perturbative final states
($gg$, $ggg$ and $\qqbar g$)
have been studied in detail using perturbation theory
\cite{Kwong,Kopke,Schuler,Braaten},
and appear to be fairly well understood.
It is therefore improbable that the treatment of the $\ccbar$ binding
should require large corrections.
This conclusion is supported by the fact that
the relative rate of $\psi'$ and direct
$\jp$ production (Eq.~(\ref{psiprime})),
which at high energies should be independent of the production mechanism,
is in agreement with experiment.

In a leading twist description, an incorrect normalization of the
charmonium production cross sections can arise from large higher order
corrections or uncertainties in the parton distributions \cite{Schuler}.
Even if the normalization is wrong by as much as a factor
of 10, such
a $K$-factor would not explain the $\jp$ polarization
data. Thus a
more likely explanation of the discrepancy may be that there are
important higher-twist contributions
to the production of the $\jp$, $\psi'$
and $\chi_1$.

The direct $\jp$ and $\chi_1$ subprocesses require,
at leading order and twist,
the emission of a quark or gluon, \eg, $gg \to \jp+g$. This
implies a higher subenergy $\sqrt{\widehat{s}}$
for these processes compared to that for the $\chi_2$, which can
be produced through simple gluon fusion, $gg \to \chi_2$.
It is then plausible
that a higher twist component which
avoids the necessity for
gluon emission is more significant for the $\jp$ and the
$\chi_1$ than it is for the $\chi_2$
(or for lepton pair production, $\qqbar
\to \gamma^*$). If either the projectile
or the target contributes {\em two}
partons rather than one, no emission of a parton is required
\cite{EE}: $(gg)+g\to \jp$. Similar production mechanisms are
considered in Ref. \cite{BHMT}.

Taking two partons from the same hadron
is a higher twist process, and as such
is suppressed by a factor of $\order(\Lambda_{QCD}^2/m_c^2)$. This factor
describes the probability that the two
partons are within a transverse distance
of $\order(1/m_c)$, as required if both of them are to couple to the same
$\ccbar$ pair. For the $\jp$ and $\chi_1$, this suppression is
compensated by the fact that the subprocess
energy $\widehat s$ can be equal
to the charmonium mass since
no parton needs to be emitted. Finding two softer
gluons in a hadron may also be more probable than the probability for one
gluon carrying the full momentum.

In the $x_F \to 1$ limit, important higher twist effects are expected
\cite{BerBro,BBKM,Eskola} and observed
\cite{Heinrich} also in the muon pair
production process, $\pi N \to \mu^+\mu^- +X$.
In effect, both valence quarks
in the pion projectile must be involved in the
reaction if the full momentum is
to be delivered to the muons. The higher twist
effect manifests itself in the
angular distribution of the muons: the polarization of the virtual photon
changes from transverse to longitudinal at large $x_F$.
Thus the photon tends
to carry the same helicity as the pion in the
$x_F \to 1$ limit. It is natural
to expect the higher twist effects to be
similarly enhanced in $\jp$ production
at large $x_F$. As seen in Fig. 1b, the
data does indeed show a remarkable
turnover in the polarization of the $\jp$
for $x_F\gsim 0.8$, with the fastest
$\jp$'s being longitudinally polarized. In contrast to the lepton pair
production case, the evidence for higher
twist effects persists, as we have
seen, for $\jp$'s produced even at lower momentum fractions.

It has recently been pointed out \cite{BDF,RS} that there is also a large
discrepancy between the leading twist QCD prediction and data for
large $p_{\perp}$ charmonium production. At leading twist in
$1/p_{\perp}^2$ and in $1/m_{c}^2$ the dominant source of ``prompt"
$\jp$'s (\ie, those not due to $B \rightarrow \jp + X$ decays) is
predicted to be the radiative decays of $P$-wave charmonia,  $\chi_{c}
\rightarrow \jp + \gamma$. The $\jp$ cross section obtained this way is
consistent with the data within a factor $\sim 2$. The actual
$\chi_{c}$ production cross section has not yet been measured. However,
the prediction for $\psi'$ production (which cannot be produced via
radiative decays) is too low by a factor $\sim 30$ when compared to the
data. At the same time, the experimental ratio of the $\psi'$ and total
(prompt) $\jp$ cross section is consistent with the universal ratio of
Eq. (\ref{psiprime}). This suggests that a major part of the prompt high
$p_{\perp}$ $\jp$'s are produced directly, rather than via  $\chi_{c}
\rightarrow \jp + \gamma$ decays. Since the shape of the
$p_{\perp}$-distribution of the measured $\jp$ and $\psi'$ cross
sections is in agreement with the leading twist prediction \cite{BDF,RS},
the large higher twist corrections are likely to reside in the $g
\rightarrow \jp$ and $c \rightarrow \jp$ fragmentation vertices, and
thus be of ${\cal O} (1/m_{c}^2)$ rather than of ${\cal
O}(1/p_{\perp}^2$). This is consistent with our conclusions based on
the low $p_{\perp}$ charmonium data. The much larger discrepancy at
high $p_{\perp}$ is qualitatively expected since the high $z$ region of
the fragmentation is emphasized due to the ``trigger bias" effect. As
discussed above, the higher twist ``intrinsic charm" mechanism is
particularly important at high momentum fraction in either the projectile
or fragmenting  parton systems.

Additional
independent evidence for higher twist effects in $\jp$ production
is also reflected in
the nuclear target $A$-dependence of the cross section.
In lepton pair production, the
cross section is very closely linearly
dependent on $A$ (apart from a small
deviation at the largest $x_F$ \cite{AldeDY}).
$\jp$ production, on the other
hand, shows a nuclear suppression over the
whole $x_F$ range \cite{AldeJP}. The
suppression depends on $x_F$ rather than on
$x_2$, and it is thus possible to
conclude \cite{HVS} that QCD factorization
must be broken, implying that the
effect is due to higher twist terms.

Further theoretical work is needed to establish
that the data on direct $\jp$
and $\chi_1$ production indeed can be described
using a higher twist mechanism
of the type discussed here. Experimentally,
it is important to check whether the
$\jp$'s produced indirectly via $\chi_2$ decay
are transversely polarized. This
would show that $\chi_2$ production is
dominantly leading twist, as we have
argued. Better data on real or virtual photoproduction of the
individual
charmonium states would also add important
information. So far, little is known
about the relative size of direct and indirect
$\jp$ photoproduction, and the
polarization measurements
\cite{ClarkPRL45,AubertNPB213} are too inaccurate to
test theoretical predictions
\cite{Berger,BaierNPB201,BaierNPB218,Korner}.
\goodbreak

In photoproduction one expects less
higher twist effects associated with the
projectile,
since in the case of direct photon interactions
only a single beam parton (the photon itself) is available. However,
the target hadron can contribute two gluons.
In the special case of diffractive
$\jp$ photoproduction \cite{Sokoloff}, this is in fact expected to be the
dominant reaction mechanism \cite{Ryskin}.

\begin{flushleft}
{\bf Acknowledgements} \\
PH is grateful for the warm hospitality of the SLAC theory division
during a visit when this work was completed.
\end{flushleft}

\bigskip

\appendix

\section{Density matrix elements}

The density matrix elements for $\jp$ production in the $gg \to \jp + g$
subprocess are defined as
\begin{equation}
  \varrho_{\mu\mu'} \equiv \frac{1}{256} \hspace{-3mm}
  \sum_{\begin{array}{c} \mu_1 \mu_2 \mu_3 \\ c_1 c_2 c_3 \end{array}}
  \hspace{-5mm}
  A(g_1 g_2 \to \jp(\mu) +  g_3) A^*(g_1 g_2 \to \jp(\mu') +  g_3),
\end{equation}
where $\mu_i, c_i$ are the helicity and color of the gluon $i$, and the
factor of 1/256 comes
from averaging over the
initial helicities and colors.
The diagonal matrix elements are found to be
\begin{eqnarray}
  \rhII & = & \frac{40 \pi^2 \als^3 \RS M}{9 [(s-M^2)(t-M^2)(u-M^2)]^2}
    \nonumber \\ & & \times
    \left\{ s^2(s-M^2)^2 + t^2(t-M^2)^2 + u^2(u-M^2)^2
    \right. \nonumber \\ & & \left. \mbox{}
    - 2M^2 [ \kbt^2 (s^2+t^2) + 2\kbt\cdot\ktt s^2 + \ktt^2 (s^2+u^2) ]
    \right\},
    \label{rho11dir} \\
\nonumber \\
  \rhOO & = & \frac{40 \pi^2 \als^3 \RS M}{9 [(s-M^2)(t-M^2)(u-M^2)]^2}
    \nonumber \\ & & \times
    \left\{  s^2(s-M^2)^2 + t^2(t-M^2)^2 + u^2(u-M^2)^2
    \right. \nonumber \\ & & \left. \mbox{}
    - 4M^2 [ \kbz^2 (s^2+t^2) + 2\kbz\ktz s^2 + \ktz^2 (s^2+u^2) ]
    \right\},
    \label{rho00dir}
\end{eqnarray}
where $M$ is the $\jp$ mass, $s,t,u$ are the subprocess invariants
(the carets being omitted for clarity), and ${\bf k}_{1,2}$ are the
three-momenta of the beam and target
partons in the Gottfried--Jackson frame.

In analogous notation, the diagonal density matrix elements for
$\chi_1$ production in
$\qqbar$, $gq$ and $gg$
scattering are
\begin{eqnarray}
   \rhII^{\qqbar}& = & \frac{(64\pi)^2 \als^3 \RP}{9 M (s-M^2)^4}
      \left[ -t\kbt^2 + (s-M^2)\kbt\cdot\ktt - u\ktt^2 + tu \right],
      \label{rho11qqbar} \\
   \rhOO^{\qqbar}  & = & \frac{(64\pi)^2 \als^3 \RP}{9 M (s-M^2)^4}
      \left[ -2t\kbz^2 + 2(s-M^2)\kbz\ktz - 2u\ktz^2 + tu \right],
            \label{rho00qqbar} \\
   \rhII^{gq} & = & \frac{-3(64\pi)^2 \als^3 \RP}{72 M (t-M^2)^4}
      \left[ -s\kbt^2 + (u-s)\kbt\cdot\ktt + su \right],
      \label{rho00qg} \\
   \rhOO^{gq} & = & \frac{-3(64\pi)^2 \als^3 \RP}{72 M (t-M^2)^4}
      \left[ -2s\kbz^2 + 2(u-s)\kbz\ktz + su \right],
      \label{rho11qg} \\
   \rhII^{gg} & = & \frac{96 \pi^2 \als^3 \RP}{M^3 (Q- M^2 P)^4}
      \nonumber \\ & & \times
      P^2 \left[ M^2 P^2 (M^4 - 4P) - 2Q(M^8 - 5M^4 P - P^2) - 15 M^2 Q^2
      \right]
      \nonumber \\ & & \mbox{} - \half\rhOO^{gg},
      \label{rho11gg} \\
\nonumber \\
   \rhOO^{gg} & = & \frac{48 \pi^2 \als^3 \RP}
   {M stu [(s-M^2)(t-M^2)(u-M^2)]^4}
      \nonumber \\
      & & \times \left\{ s^2(s-M^2)^2 [\kbz g(s,t,u) + \ktz g(s,u,t)]^2
      \right. \nonumber \\
      & & \mbox{} + u^2(u-M^2)^2 [\kbz g(u,t,s) - (\kbz+\ktz) g(u,s,t)]^2
      \nonumber \\
      & & \mbox{} + t^2(t-M^2)^2 [(\kbz+\ktz) g(t,s,u) -\ktz g(t,u,s) ]^2
      \nonumber \\
      & & \mbox{} + 4M^2 (k_{1x} k_{2y} - k_{2x} k_{1y})^2
      \nonumber \\ & & \times \left.
      \left[ s^2(s-M^2)^2 f^2(s,t,u) + (s \lra u) + (s \lra t) \right]
      \right\}. \label{rh00gg}
\end{eqnarray}
The density matrix elements ${\varrho}^{qg}$
for the processes where a beam
quark scatters off a target gluon are obtained by changing
${\bf k}_1 \lra {\bf k}_2$
(and consequently $t \lra u$) in ${\varrho}^{gq}$.
The matrix elements for the
$g\bar{q}$ and $\bar{q}g$ scattering processes are
the same as for $gq$ and $qg$, respectively.
In deriving ${\varrho}^{gg}$
we made use of the subprocess amplitudes given
in Ref. \cite{GastmansWu}.
The functions $f,g,P$ and $Q$ of the invariants
are defined as
\begin{eqnarray}
  f(s,t,u) & = & (t-u)(st+tu+us-s^2), \\
  g(s,t,u) & = & (s+t) [ st(t-s) + su(u-s) + tu(t-u) ], \\
  P & = & st+tu+us, \\
  Q & = & stu.
\end{eqnarray}

\bigskip
\centerline{\large FIGURE CAPTION}
\vspace{1cm}
{\bf Figure 1.} Leading-twist predictions
of the parameter $\lambda$ in the
decay angular distribution of $\jp$'s produced
in pion-nucleon collisions at
$E_{\rm lab}=300$ GeV, plotted as a function of $x_F$.
{\em (a)} The three solid curves show the decay distributions of  $\jp$'s
produced via radiative decays of the $\chi_2$ and $\chi_1$ states and
``directly" in gluon fusion. The dashed curves show the effect
of smearing the transverse momentum distribution of the beam
parton by a Gaussian function
$\exp \left[ -(k_\perp/500 \; {\rm MeV})^2 \right] $.
{\em (b)} The combined decay distribution of all $\jp$'s, including
contributions from $\chi_{1,2}$ decays and
direct production, is shown here.
The lower curve shows the effect of adjusting the relative
normalization of the different
contributions to their measured values (see
Table 2) by appropriate $K$-factors. The dashed curve shows the effect of
transverse momentum smearing and $K$-factors adjustments .
The data is from the Chicago--Iowa--Princeton (252-GeV $\pi$W
collisions, Ref. \cite{Biino};
full circles)
and E537 (125-GeV $\pi$W collisions,
Ref. \cite{Akerlof};
open circles)
experiments.

\end{document}